\documentstyle[12pt,aps,prd,preprint]{revtex}
\begin{document}
\title{Late-Time Evolution of Charged Gravitational Collapse and 
Decay of Charged Scalar Hair - I}
\author{Shahar Hod and Tsvi Piran}
\address{The Racah Institute for Physics, The
Hebrew University, Jerusalem 91904, Israel}
\date{\today}
\maketitle

\begin{abstract}
We study {\it analytically} the asymptotic evolution of {\it charged} fields
around a Reissner-Nordstr\"om black-hole. Following the no-hair theorem we
focus attention on the {\it dynamical} mechanism by which the charged hair is
radiated away. We find an inverse power-law relaxation of the charged fields at
future timelike infinity, along future null infinity and an oscillatory
inverse power-law relaxation along the future outer horizon. We show that a
charged hair is shedd {\it slower} than a neutral one. Our results are also of
importance to the study of mass inflation and the stability of Cauchy
horizons during a dynamical gravitational collapse of charged matter to form a
charged black-hole.
\end{abstract}

\section{Introduction}\label{Introduction}
The statement that black-holes have no-hair was introduced by Wheeler in the
early 1970s. The various no hair theorems state that the external field of a
black-hole relaxes to a Kerr-Newman field described solely by three parameters:
the black-hole's mass, charge and angular-momentum. The mechanism responsible for
the relaxation of {\it neutral} external perturbations was first studied by
Price \cite{Price}. He has found that the late-time behaviour of these
neutral perturbations for a fixed position, $r$, is dominated by $t^{-(2l+3)}$
tails (if there is no initial static field), where $l$ is the
multipole moment of the mode and $t$ is the standard Schwarzschild time
coordinate. The behaviour of {\it neutral} perturbations along null infinity and
along the future event horizon was further studied by Gundlach, Price and Pullin
\cite{Gundlach}. They have found that the neutral perturbations along null
infinity decay according to an inverse power-law $u^{-(l + 2)}$, where $u$ is
the outgoing Eddington-Finkelstein null coordinate. Along the event-horizon the
perturbations decay according to $v^{-(2l + 3)}$, where $v$ is the
ingoing Eddington-Finkelstein null coordinate.

In this work we study the gravitational collapse of a {\it charged} matter to
form a charged black hole. In such a collapse one should expect that {\it
charged} perturbations will develop outside the collapsing star. In particular
we focus attention on the late-time behaviour of such {\it charged} scalar
perturbations along these three asymptotic regions.

Our results are of importance for two major areas of black-hole physics:

\begin{enumerate}
\item The no-hair theorem of Mayo \& Bekenstein \cite{MayoBek} states that
black-holes cannot have a charged scalar-hair. However, it was never before
studied how a charged black-hole, which is formed during a gravitational
collapse of a {\it charged} matter, dynamically sheds its charged scalar hair
during the collapse. We study, here, the mechanism by which the
{\it charged} hair is radiated away.

\item The mass-inflation scenario and the stability of Cauchy horizons were
studied under the assumption of the existence of inverse power-law (neutral)
perturbations along the outer horizon of a Reissner-Nordstr\"om black-hole.
However, these models did not take into account the existence of {\it charged}
perturbations which are expected to appear in the dynamical collapse of a
{\it charged} star. Here, we study the asymptotic behaviour of such
perturbations.
\end{enumerate}

The plan of the paper is as follows. In Sec. \ref{description} we describe our
physical system and formulate the evolution equation. In Sec.
\ref{evolutioncollapse} we study the late-time evolution of charged scalar
perturbations for a collapse that leads to the formation of a
(charged) black-hole. Here we generalize the formalism of Ref.
\cite{Price,Gundlach} to the charged situation. We study the case $|eQ| \ll
1$, which simplifies things enough to allow us {\it analytical} derivations of
our results. In paper II in this series we will examine the problem for a
general value of $|eQ|$. We find an inverse power-law behaviour of the
charged perturbations along the three asymptotic regions. However, the exponents
differ for those of neutral perturbations. Additionally along the outer horizon
there are periodic {\it oscillations} on top of this power law decay (which do
not exist for neutral perturbations).

In Sec. \ref{evolutionnon-collapse} we study the behaviour of charged
perturbations in the non-collapsing case (imploding and exploding shells).
Qualitatively, we find the same late-time behaviour as in the collapsing
situation. In Sec. \ref{hair} we compare the late-time behaviour of charged
perturbations with the late-time behaviour of neutral perturbations. We find
that the dynamical process of shedding hair is different for neutral hair and
charged one, both quantitatively and qualitatively. We show that a black-hole
which is formed from the gravitational collapse of a {\it charged} matter
becomes ``bald" {\it slower} than a neutral one due to the existence of charged
perturbations. Furthermore, while the late-time behaviour of neutral
perturbations is determined by the space-time curvature, the late-time behaviour
of charged fields is dominated by {\it flat} space-time effects (scattering due
to the {\it electromagnetic} interaction in {\it flat} space-time). We conclude
in Sec. \ref{summary} with a brief summary of our results.

\section{Description of the system}\label{description}

The external gravitational field of a spherically symmetric collapsing star of
mass $M$ and charge $Q$ is given by the Reissner-Nordstr\"om metric

\begin{equation}\label{RNmetric}
ds^2=-\left( {1-{{2M} \over r}+{{Q^2} \over {r^2}}} \right)dt^2+\left( {1-{{2M}
\over r}+{{Q^2} \over {r^2}}} \right)^{-1}dr^2+r^2d\Omega ^2\  .
\end{equation}

We will also use the tortoise radial coordinate $y$, defined by $dy={{dr}
\mathord{\left/ {\vphantom {{dr} {\left( {1-{{2M} \over r}+{{Q^2} \over {r^2}}}
\right)}}} \right. \kern-\nulldelimiterspace} {\left( {1-{{2M} \over r}+{{Q^2}
\over {r^2}}} \right)}}$, in terms of which the metric becomes

\begin{equation}\label{RNmetricy}
ds^2=\left( {1-{{2M} \over r}+{{Q^2} \over {r^2}}} \right)\left( {-dt^2+dy^2}
\right)+r^2d\Omega ^2\  ,
\end{equation}
where $r=r(y)$.

We consider the evolution of massless {\it charged} scalar perturbations fields
outside a charged collapsing star. The wave equation for the complex
scalar-field is \cite{HawkingEllis}

\begin{equation}\label{varyingphi}
\phi _{;ab}g^{ab}-ieA_ag^{ab}\left( {2\phi _{;b}-ieA_b\phi }
\right)-ieA_{a;b}g^{ab}\phi =0\  ,
\end{equation}
where $e$ is constant.

Resolving the charged scalar-field into spherical harmonics $\phi
=\sum\limits_{l,m} {\eta _m^l\left( {t,r} \right)Y_l^m{{\left( {\theta ,\varphi
} \right)} \mathord{\left/ {\vphantom {{\left( {\theta ,\varphi } \right)} r}}
\right. \kern-\nulldelimiterspace} r}}$, we obtain a wave equation for each
multipole moment

\begin{equation}\label{wavemult}
\eta _{,tt}-2ieA_t\eta _{,t}-\eta _{,yy}+\tilde V\eta =0\  ,
\end{equation}
where
\begin{equation}\label{Vtilde}
\tilde V=\tilde V_{M,Q,l,e}\left( r \right)=\left( {1-{{2M} \over r}+{{Q^2}
\over {r^2}}} \right)\left[ {{{l\left( {l+1} \right)} \over {r^2}}+{{2M} \over
{r^3}}-{{2Q^2} \over {r^4}}} \right]-e^2A_t^2\  .
\end{equation}
Here we have suppressed the indices $l,m$ on $\eta$.

The electromagnetic potential satisfies the relation

\begin{equation}\label{empot}
A_t=\Phi -{Q \over r}\ .
\end{equation}
where $\Phi$ is constant.

In order to get rid of the, physically unimportant quantity $\Phi$, we introduce
the auxiliary field $\psi =e^{-ie\Phi t}\eta $, in terms of which the equation of
motion (\ref{wavemult}) becomes

\begin{equation}\label{neweqmotion}
\psi _{,tt}+2ie{Q \over r}\psi _{,t}-\psi _{,yy}+V\psi =0\  ,
\end{equation}
where

\begin{equation}\label{defV}
V=V_{M,Q,l,e}\left( r \right)=\left( {1-{{2M} \over r}+{{Q^2} \over {r^2}}}
\right)\left[ {{{l\left( {l+1} \right)} \over {r^2}}+{{2M} \over {r^3}}-{{2Q^2}
\over {r^4}}} \right]-e^2{{Q^2} \over {r^2}}\  .
\end{equation}

It is well known that a gauge transformation of the form $\eta \to e^{-i \alpha
t} \eta$ (where $\alpha$ is a real constant) merely adds a constant to $A_t$,
i.e. $A_t \to A_t - {1 \over e} \alpha$.

\section{EVOLUTION OF CHARGED PERTURBATIONS IN THE COLLAPSING CASE (BLACK-HOLE
FORMATION)}\label{evolutioncollapse}
The general solution to the wave-equation (\ref{neweqmotion}) can be written as

\begin{eqnarray}\label{solweq}
\psi = & \sum\limits_{k=0}^l & {A_kr^{-k}\left[ {e^{-ieQ\ln r}G^{\left( {l-k}
\right)}\left( u \right)+\left( {-1} \right)^ke^{ieQ\ln r}F^{\left( {l-k}
\right)}\left( v \right)} \right]}\nonumber \\
+ & \sum\limits_{k=0}^\infty & {\left[ {B_k\left(
r \right)G^{\left( {l-k-1} \right)}\left( u \right)+C_k\left( r \right)F^{\left(
{l-k-1} \right)}\left( v \right)} \right]}\  ,
\end{eqnarray}
where $G$ and $F$ are arbitrary functions. The coefficients $A_k = A_k(l) =
(l+k)!/[2^kk!(l-k)!]$ are equal to those that arise in the neutral case
\cite{Price} and $B_k(r) = B_k(r;eQ,l,M), C_k(r) = C_k (r;eQ,l,M)$. Here $u
\equiv t - y$ is a retarded time coordinate and $v \equiv t + y$ is an advanced
time coordinate. For any function $H, H^{(k)}$ is the $k -
th$ derivative of $H^{(0)}$; negative-order derivatives are to be interpreted as
integrals. The first sum in (\ref{solweq}) represents the zeroth-order solution,
i.e. neglecting terms of order $O(eQ), O({M \over r}), O({Q \over r})$ and
higher.

The functions $B_k(r)$ satisfy the recursion relation

\begin{eqnarray}\label{recursion}
&2\lambda ^2B'_k+2ieQB_kr^{-1}-\lambda ^2\left( {B'_{k-1}\lambda ^2}
\right)'-\lambda ^2\left[ {A_k\left( {-k-ieQ} \right)r^{-k-1-ieQ}\lambda
^2} \right]'-\nonumber \\
&2A_{k+1}r^{-k-2-ieQ}\left[ {ieQ\left( {\lambda ^2-1} \right)+\lambda ^2\left(
{k+1} \right)} \right]+V\left( r \right)\left[ {A_kr^{-k-ieQ}+B_{k-1}}
\right]=0\  ,
\end{eqnarray}
where $\lambda ^2\equiv 1-{{2M} \over r}+{{Q^2} \over {r^2}}$ and $B'\equiv {dB
/dr}$.

The functions $C_k(r)$ satisfy an analogous recursion relation; however, we
will not need them as the late-time behaviour of the charged scalar-field does
not depend on $C_k$. We can now expand $B_k(r)$ in the form

\begin{equation}\label{expandBk}
B_k\left( r \right)=a_kr^{-\left( {k+1} \right)-ieQ}+b_kr^{-\left( {k+2}
\right)-ieQ}+\cdots\  ,
\end{equation}
where $a_k\equiv a_k(l,eQ), b_k \equiv b_k(l,eQ) \cdots$.

Substituting (\ref{expandBk}) in (\ref{recursion}), one finds for the lowest
order coefficients

\begin{equation}\label{defal}
a_l=-ieQA_l{{2l+1} \over {2\left( {l+1} \right)}}\left[ {1+O\left( {eQ} \right)}
\right]\  .
\end{equation}

The star begins to collapse at retarded time $u=u_0$. The world line of the
stellar surface is asymptotic to an ingoing null line $v=v_0$, while the
variation of the field on the stellar surface is asymptotically infinitely
redshifted \cite{Price,Bicak}. This effect is caused by the time dilation
between static frames and infalling frames. A static external observer sees all
processes on the stellar surface become ``frozen" as the star approaches the
horizon. Thus, he sees all physical quantities approach a constant. Using the
above effect, we make the explicit assumption that after some retarded time
$u=u_1, D_u\phi=0$ on $v=v_0$, where $D_\mu=\partial_\mu - ieA_\mu$ is the
gauge covariant derivative. This assumption has been proven to be very successful
for the neutral scalar-field \cite{Price,Gundlach}.

We start with the first stage of the evolution, i.e. the scattering of the
charged field in the region $u_0 \leq u \leq u_1$:

The first sum in (\ref{solweq}) represents the primary waves in the wave front,
while the second sum represents backscattered waves. The interpretation of
these integral terms as backscatter comes from the fact that they depend on
data spread out over a {\it section} of the past light cone, while outgoing
waves depend only on data at a fixed $u$ \cite{Price}. It should be noted that
this physical distinction between the primary waves and the backscattered waves
is valid in the region $r \ll |Q| e^{{1}\over{|eQ|}}$.

After the passage of the primary waves there is no outgoing radiation for $u >
u_1$, aside from backscattered waves. This means that $G(u_1) = 0$. Thus, for a
large $r$ at $u = u_1$, the dominant term in (\ref{solweq}) is

\begin{equation}\label{domterm}
\psi \left( {u=u_{1,}r} \right)=a_lr^{-\left( {l+1} \right)}G^{\left( {-1}
\right)}\left( {u_1} \right)\  .
\end{equation}
This is the dominant backscatter of the primary waves.

After we had determined the dominant backscatter of the charged scalar field,
we shall next consider the asymptotic evolution of the field. We confine our
attention to the region $y \gg M,|Q|; u>u_1$. In this region (and for $|eQ| \ll
1$) the evolution of the field is dominated by {\it neutral flat} space-time
terms, i.e.

\begin{equation}\label{flatst}
\psi _{,tt}-\psi _{,rr}+{{l\left( {l+1}
\right)} \over {r^2}}\psi =0\  .
\end{equation}
[It should be noted that for $r \gg M,|Q|$, we have $\psi _{,yy} \simeq \psi
_{,rr}+{{2M} \over {r^2}}\psi _{,r}$. However, in this region $O\left( {\psi
_{,rr}} \right)=O\left( {r^{-2}\psi } \right) \gg O\left( {Mr^{-2}\psi _{,r}}
\right)$. So, in this region, we may replace $r$ by $y$.]

Thus, the solution for $\psi$ can be written as

\begin{equation}\label{solpsi}
\psi =\sum\limits_{k=0}^l {A_ky^{-k}\left[ {g^{\left( {l-k} \right)}\left( u
\right)+\left( {-1} \right)^k f^{\left( {l-k} \right)}\left( v \right)}
\right]}\  .
\end{equation}
Comparing (\ref{solpsi}) with the initial data (\ref{domterm}) on $u=u_1$, one
finds

\begin{equation}\label{comparepsi}
f\left( v \right)=F_0v^{-1}\  ,
\end{equation}
where

\begin{equation}\label{defefzero}
F_0={{ieQG^{\left( {-1} \right)}\left( {u_1} \right)\left( {-1}
\right)^{l+1}\left( {2l+1} \right)!!} \mathord{\left/ {\vphantom {{ieQG^{\left(
{-1} \right)}\left( {u_1} \right)\left( {-1} \right)^{l+1}\left( {2l+1}
\right)!!} {\left( {l+1} \right)!+O\left[ {\left( {eQ} \right)^2} \right]}}}
\right. \kern-\nulldelimiterspace} {\left( {l+1} \right)!+O\left[ {\left( {eQ}
\right)^2} \right]}}\  .
\end{equation}
For late times $t \gg y$ we can expand $g\left( u
\right)=\sum\limits_{n=0}^\infty  {{{\left( {-1} \right)^n} \over {n!}}g^{\left(
n \right)}\left( t \right)y^n}$ and $f\left( v \right)=\sum\limits_{n=0}^\infty
{{1 \over {n!}}f^{\left( n \right)}\left( t \right)y^n}$. Using these
expansions we can rewrite (\ref{solpsi}) as

\begin{equation}\label{againsolpsi}
\psi =\sum\limits_{n=-l}^\infty  {K_ny^n\left[ {f^{\left( {l+n}
\right)}\left( t \right)+\left( {-1} \right)^ng^{\left( {l+n}
\right)}\left( t \right)} \right]}\  ,
\end{equation}
where the coefficients $K_n$ are those given in the neutral case \cite{Price}.

Using the boundary conditions for small $r$ (regularity as $y \to -\infty$, at
the horizon of a black-hole, or at $r=0$, as in the next section), one finds that
at late times the terms $h\left( t \right)\equiv f\left( t \right)+\left( {-1}
\right)^lg\left( t \right)$ and $f^{(2l+1)}(t)$ must be of the
same order (see \cite{Gundlach} for additional details). Thus, we conclude that

\begin{equation}\label{defft}
f\left( t \right) \simeq F_0t^{-1}\  ,
\end{equation}

\begin{equation}\label{defgt}
g\left( t \right)
\simeq \left( {-1} \right)^{l+1}F_0t^{-1}
\end{equation}
and

\begin{equation}\label{defht}
h\left( t \right)=O\left[ {t^{-2\left( {l+1} \right)}} \right]\  .
\end{equation}
Hence, we find that the late-time behaviour of the charged scalar-field for
$|Q|e^{{{1}\over{|eQ|}}}\gg t \gg y \gg M,|Q|$ is

\begin{equation}\label{scalfieldbeh}
\psi \simeq 2K_{l+1}y^{l+1}f^{\left( {2l+1} \right)}\left(
t \right)=-2K_{l+1}F_0\left( {2l+1} \right)!t^{-2\left( {l+1}
\right)}y^{l+1}+O\left[ {\left( {eQ} \right)^2} \right]\  .
\end{equation}
This is the late-time behaviour of the charged scalar field at timelike
infinity $i_+$.

>From equations (\ref{solpsi}), (\ref{comparepsi}) and (\ref{defgt}) one finds that the
behaviour of the charged scalar field at future null infinity $scri_+$ (i.e., at
$u \ll v \ll |Q|e^{{{1}\over{|eQ|}}}$) is

\begin{equation}\label{scalfieldbehvinfty}
\psi \left( {v \gg u,u} \right) \simeq A_0g^{\left( l \right)}\left(
u \right) \simeq -F_0l!u^{-(l+1)}\  .
\end{equation}
Finally, we go on to consider the behaviour of the charged scalar-field at the
black-hole outer horizon $r_+$:

As $y \to - \infty$ the wave-equation (\ref{neweqmotion}) can be approximated by
the equation

\begin{equation}\label{waveminusinfty}
\psi _{,tt}+2ie{Q \over {r_+}}\psi _{,t}-\psi _{,yy}-{e^2Q^2 \over {r^2_+}}\psi
=0\  ,
\end{equation}
whose general solution is

\begin{equation}\label{gensolwave}
\psi =e^{-ie{\textstyle{Q \over {r_+}}}t}\left[ {\alpha
\left( u \right)+\gamma \left( v \right)} \right]\  .
\end{equation}

On $v = v_0$ we take $D_u\phi=0$ (for $u \to \infty$). Thus, $\alpha(u)$
must be a constant, and with no loss of generality we can choose it to be zero.
Next, we expand $\gamma(v)$ for $t \gg |y|$ as

\begin{equation}\label{phigammaextended}
\psi =e^{-ie{\textstyle{Q \over {r_+}}}t}\gamma \left( v
\right)=e^{-ie{\textstyle{Q \over {r_+}}}t}\sum\limits_{n=0}^\infty  {{1 \over
{n!}}\gamma ^{\left( n \right)}\left( t \right)y^n}\  .
\end{equation}
In order to match the $y \ll -M$ solution (\ref{phigammaextended}) with the $y
\gg M$ solution (\ref{scalfieldbeh}), we make the ansatz $\psi \simeq \psi
_{stat}\left( r \right)t^{-2\left( {l+1} \right)}$ for the solution
in the region $y \ll -M$ and $t \gg |y|$. In other words, we assume that the
solution in the $y \ll -M$ region has the same late-time $t$-dependence as the
$y \gg M$ solution. Using this assumption, one finds $\psi _{stat}=\Gamma
_0e^{ie{\textstyle{Q \over {r_+}}}y}$ and $\gamma \left( t \right)=\Gamma
_0e^{ie{\textstyle{Q \over {r_+}}}t}t^{-2\left( {l+1} \right)}$
(where $\Gamma _0$ is a constant). Thus, the late-time behaviour of the charged
scalar-field at the horizon is (for $v \ll |Q|e^{{{1}\over{|eQ|}}}$)

\begin{equation}\label{latetimescalfieldhor}
\psi \left( {u\to \infty ,v} \right)=\Gamma
_0e^{ie{\textstyle{Q \over {r_+}}}y}v^{-2\left( {l+1}\right)}\  .
\end{equation}

\section{Evolution of charged perturbations in the non-collapsing case (imploding
and exploding shells)}\label{evolutionnon-collapse}

We now consider the case of imploding and exploding shells of charged scalar
field. In this situation, the charge of the space-time falls to zero as the
evolution proceeds (the charged scalar field, which is the source of that
charge, escapes to infinity). One might think that this dissipation of the
charge of the spacetime would lead to a late-time evolution which is identical
with the neutral case. However, this is {\it not} the case; as we have seen,
the late-time behaviour of the charged scalar field at constant $r$ and at
$scri_+$ depends on the {\it backscattering} of the initially outgoing waves.
This backscattering is different for the charged scalar field compared with
the neutral one, and hence would lead to a different late-time behaviour.
It should be noted that the backscattering is taking place in an early stage
$(u<u_1)$, when the spacetime still contains a considerable amount of charge.
Thus, the initial data on $u = u_1$ is still given by (\ref{domterm}).
Furthermore, the late-time behaviour of the charged scalar field at constant $r$
and at $scri_+$ are independent of the small-$r$ nature of the background (this
is the situation in the neutral case as well \cite{Price,Gundlach}). Thus, we
expect that the non-collapsing charged field would produce a similar late-time
behaviour compared with the collapsing one.

Finally, it should be noted that in this situation the space-time has no internal
``infinity" (no event horizon) and thus $r_+ = \infty$.

\section{Charged scalar-hair vs. neutral scalar hair}\label{hair}

The no-hair theorems for black-holes state that black-hole can have neither
neutral scalar hair \cite{Chase,Bekenstein} nor a charged scalar-field hair
\cite{MayoBek}. Price \cite{Price} has investigated the mechanism which leads to
the relaxation of such external neutral scalar hair. However, it was never
before investigated how a charged black-hole, which is formed during a
gravitational collapse of a charged matter, dynamically sheds its {\it charged}
scalar hair during the process of the collapse.

Here, we have shown that the two mechanisms are quite different, both
quantitatively and qualitatively. While the late-time behaviour of a neutral
scalar field outside a black-hole is dominated by $\sim t^{-\left( {3+2l}
\right)}$ tails, the relaxation of the {\it charged} scalar hair outside a
charged black-hole is dominated by a $\sim t^{-\left({2+2l} \right)}$ behaviour.
Therefore, we conclude that charged perturbations die {\it slower} than neutral
ones, i.e. a charged black-hole, which is formed during a gravitational collapse
of a charged matter, is expected to loose its {\it charged} scalar-hair and
relax to its final state {\it slower} than a neutral one. In a more pictorial
way, a black-hole, which is formed from the gravitational collapse of a {\it
charged} matter, becomes ``bald" slower than a neutral one.

Mathematically, it is the relation of $r$ to $y$ which determines the dominant
initial backscattering (and therefore the behaviour of the late-time tails) for
{\it neutral} perturbations \cite{Price}. This means that to a leading order in
$M$ the evolution of neutral perturbations depends on the space-time curvature
(on $M$) in the first step $(u_0 < u < u_1)$, but not in the second step.

On the other hand, the {\it flat} space-time {\it charged} terms in the
evolution equation for the charged scalar-field are of critical importance in
determining the dominant initial backscattering for {\it charged}
perturbations. Indeed, the flat space-time evolution equation is

\begin{equation}\label{flatspacetime}
\psi _{,tt}+2ie{Q \over r}\psi _{,t}-\psi _{,rr}+{{l(l+1)-e^2Q^2} \over {r^2
}}\psi=0\   ,
\end{equation}
whose general solution can be written as

\begin{equation}\label{gensolflat}
\psi =\sum\limits_{k=0}^\infty  {r^{-k}\left[ {e^{-ieQ\ln r}C_kp^{\left( {-k}
\right)}\left( u \right)+e^{ieQ\ln r}D_kq^{\left( {-k} \right)}\left( v
\right)} \right]}\   ,
\end{equation}
where

\begin{eqnarray}\label{defCkandDk}
C_k & = &C_k\left( {l,eQ} \right)={1 \over {2^kk!}}\prod\limits_{n=0}^{k-1}
{\left[ {l\left( {l+1} \right)-n\left( {n+1} \right)-ieQ\left( {2n+1} \right)}
\right]} \nonumber \\
D_k & = & D_k\left( {l,eQ} \right)=(-1)^kC^*_k\   ,
\end{eqnarray}
for $k \geq 1$ and $C_0 = D_0 \equiv 1$.

For $eQ = 0$ this infinite series is cut-off at $k = l+1$, i.e.

\begin{equation}\label{cutoff}
{{C_{l+1}} \over {C_l}}={{D_{l+1}} \over {D_l}}=-ieQ{{2l+1}\over{2(l+1)}}\  .
\end{equation}
In other words, for $eQ = 0$ there is {\it no} backscatter of the waves.

For $|eQ| \ll 1$, we may rewrite (\ref{gensolflat}) as

\begin{eqnarray}\label{newgensolflat}
\psi \simeq & \sum\limits_{k=0}^l & {A_kr^{-k}\left[ {e^{-ieQ\ln r}G^{\left(
{l-k} \right)}\left( u \right)+\left( {-1} \right)^ke^{ieQ\ln r}F^{\left( {l-k}
\right)}\left( v \right)} \right]}+ \nonumber \\
& \sum\limits_{k=l+1}^\infty &  {r^{-k}\left[ {e^{-ieQ\ln r}C_kG^{\left( {l-k}
\right)}\left( u \right)+e^{ieQ\ln r}D_kF^{\left( {l-k} \right)}\left( v
\right)} \right]}\   ,
\end{eqnarray}
where $A_k$ are the coefficients given in sec. \ref{evolutioncollapse} and $C_k
= A_k + O(eQ)$ for $0 \leq k \leq l$. Thus, we conclude that the dominant
backscatter of the primary waves, for $|Q| \ll r \ll |Q|e^{{{1}\over{|eQ|}}}$, is
given by $C_{l+1}r^{-\left( {l+1} \right)}G^{\left( {-1} \right)}\left( {u_1}
\right)$ where $C_{l+1}=-ieQA_l{{2l+1} \over {2\left( {l+1} \right)}}\left[
{1+O\left( {eQ} \right)} \right]$. This is just the result obtained earlier, see
(\ref{defal}).

The physical significance of this result is the conclusion that unlike
neutral perturbations the late-time behaviour of {\it charged} scalar-field is
entirely determined by {\it flat} space-time effects. In other words, the
scattering is caused by the {\it electromagnetic} interaction in {\it flat}
space-time.

Our results 
show that mass-inflation in a gravitational collapse 
of a charged scalar field will be 
stronger than in the collapse of a neutral fields.
Mass
inflation models have relied heavily on the existence of inverse power-law tails
along the outer horizon of a Reissner-Nordstr\"om black-hole. However, these
models did not take into account the existence of {\it charged} perturbations
outside the collapsing star (during the gravitational collapse of a charged star
to form a Reissner-Nordstr\"om black-hole we, of course, do expect to find {\it
charged} perturbations outside the star). We have investigated the behaviour of
these charged perturbations on the {\it outer} horizon of a RN black-hole. We do
find a rather similar behaviour of the charged field along the outer horizon
(see (\ref{latetimescalfieldhor})), although with a {\it different} exponent and
with periodic {\it oscillations} (which do not exist for neutral
perturbations). The {\it power-law} fall-off (times periodic oscillations) of
the {\it charged} perturbations suggests that mass-inflation should occur in
the gravitational collapse of a {\it charged} matter in which a charged
black-hole forms. Moreover, since {\it charged} perturbations have {\it smaller}
dumping exponents compared with neutral ones, they will dominate the influx
through the outer horizon and hence will be the {\it main} cause for the
mass-inflation phenomena. One should remember that the mass-function diverges
like $m(v) \approx v^{-p}e^{\kappa_0v}$ (for $v \to \infty$, near the Cauchy
horizon), where ${1 \over 2}p$ is the dumping exponent of the field
\cite{Poisson}.

\section{Summary and Conclusions}\label{summary}

We have studied the gravitational collapse of a {\it charged} matter to form a
charged black-hole. The main issue considered is the late-time behaviour of
{\it charged} scalar perturbations outside the collapsing star. We have shown
that {\it power-law} tails develop at timelike infinity (at fixed
radius at late times) and along null infinity. Along the outer horizon
there is an {\it oscillatory power-law} tail. The period of these oscillations
is determined by the quantity $eQ/r_+$. The exponents of these inverse power-law
tails are all {\it smaller} compared with neutral perturbations. Thus, we
conclude that a black-hole which is formed from the gravitational collapse of a
{\it charged} matter becomes ``bald" {\it slower} than a neutral one due to the
presence of charged perturbations.

While the late-time behaviour of neutral perturbations is determined by the
space-time curvature (mathematically, the relation between $y$ and $r$),
the asymptotic behaviour of charged fields is dominated by {\it flat} space-time
effects.

Our work reveals the {\it dynamical} mechanism by which the {\it charged}
scalar-hair is radiated away leaving behind a Reissner-Nordstr\"om black-hole.
We have shown that this mechanism differs from the neutral one both {\it
quantitatively} (different power-law exponents and oscillatory behaviour along
the black-hole outer-horizon) and {\it qualitatively} (the initial
backscattering, and thus the late-time behaviour are dominated by {\it flat}
space-time terms, namely, by the {\it electromagnetic interaction}, rather than
by curvature effects).

Furthermore, we have shown that the late-time behaviour of charged fields in the
non-collapsing situation (imploding and exploding shells) is dominated by a
similar inverse power-law behaviour both at a fixed radius (and late times) and
along null infinity.

Finally, our results are of importance for the mass-inflation scenario
and stability of Cauchy horizons. Here, we have shown that the asymptotic
behaviour of charged perturbations along the outer-horizon of a RN black-hole is
characterized by an {\it oscillatory inverse power-law} behaviour (with smaller
exponents compared with neutral tails). Thus, one should expect that these
inverse power-law {\it charged} perturbations will cause a mass-inflation
singularity during the gravitational collapse of a charged matter that forms a
{\it charged} black-hole. Moreover, the {\it slower} relaxation of charged
perturbations makes  them the {\it dominant} cause for the divergence of the
mass-function.

The most significant shortfall in our analysis is the limitation to the case
$|eQ| \ll 1$. In an accompanying paper (II) 
we extend our {\it analytical} results
to include {\it general} values of $eQ$ (using a spectral decomposition) and
we confirm them {\it numerically}. On the other hand, the main advantage of {\it
this} approach is the fact that it gives a {\it clear} picture of the {\it
physical} mechanism responsible for the late-time behaviour of charged
perturbations, namely, the tail arises because of backscattering of the {\it
charged} field off the {\it electromagnetic} potential far away from the
black-hole. The physical picture that arises from this paper is clear ---
dealing with charged (massless) perturbations, one may {\it neglect} any
curvature effects.

In accompanying papers we study the {\it fully nonlinear} gravitational collapse
of a charged scalar field to form a {\it charged} black-hole. In order to
numerically confirm our {\it analytical} predictions we will first focus
attention on the asymptotic behaviour of the {\it charged} field outside the {\it
dynamically} formed charged black-hole.

\bigskip
\noindent
{\bf ACKNOWLEDGMENTS}
\bigskip

S. Hod wishes to thank Avraham E. Mayo for helpful discussions. This research was
supported by a grant from the Israel Science Foundation.

\end{document}